\documentclass[aps,groupedaddress,showpacs]{revtex4}
\usepackage{amsmath,amscd,amssymb,epsfig}

\begin{document}
\bibliographystyle{apsrev}

\title{Quantum dynamics with non-Markovian fluctuating parameters}

\author{Igor Goychuk}
\email[]{goychuk@physik.uni-augsburg.de, goychuk@mailaps.org}

\affiliation{Institute of Physics, University of Augsburg,
Universit\"atsstr. 1, D-86135 Augsburg, Germany}

\date{\today}

\begin{abstract}

A stochastic approach to the quantum dynamics randomly modulated
in time by a discrete state  non-Markovian noise, which possesses an
arbitrary non-exponential distribution of the residence times,  
is developed. The
formally {\it exact} expression for the Laplace-transformed  quantum
propagator averaged over the {\it stationary} realizations of such
N-state non-Markovian  noise  
is obtained. 
The theory possesses a wide range of
applications. It includes  some previous  Markovian and
non-Markovian theories as particular cases. In the context of 
stochastic theory of spectral line shape and relaxation, the
developed approach presents a non-Markovian generalization of the
Kubo-Anderson theory of sudden modulation.  In particular,  
the exact analytical expression is derived for the spectral line 
shape of optical
transitions described by a Kubo-oscillator with randomly modulated 
frequency
which undergoes jump-like non-Markovian fluctuations in time. 

\end{abstract}


\maketitle

\section{Introduction}

The dynamics of classical and quantum systems in the presence of 
randomly fluctuating micro-fields of environment presents one of 
the fundamental problems in physics. Spin relaxation in solids 
\cite{anderson53,kubo}, exciton transport in molecular systems
\cite{reineker}, single-molecular spectroscopy \cite{jbs02},
classical transport processes with fluctuating barriers 
\cite{fbarrier}  present a few relevant examples.  A popular approach
consists in modeling the  ambient noise influence by means of a classical
stochastic field acting on the dynamical system. In the case of quantum 
systems, such a phenomenological approach is
known under the label of Stochastic Liouville Equation (SLE) approach
\cite{kubo63,fox,lindenberg}.  It is suitable in the limit of
sufficiently high (formally infinite) temperatures 
\cite{reineker,lindenberg}. In the field of chemical kinetics a
similar methodology is known under the label of rate processes with
dynamical disorder where the rates of chemical 
reactions fluctuate \cite{zwanzig}.
Moreover, the addition of a non-equilibrium 
classical noise into dissipative quantum dynamics can serve to describe
the influence of the non-equilibrium environmental degrees of freedom  
on the transport properties \cite{gpm95}. 

The tractability 
of the SLE approach, which allows to arrive at the exact model
solutions in several  known cases, makes 
it popular over the years. For example,
the case of two-state quantum dynamics subjected to a
white Gaussian noise 
can be treated exactly and the corresponding {\it exact} 
master equations for
the averaged parameters of quantum systems can be derived 
\cite{reineker,lindenberg}. However, already in the case of 
a colored Gaussian
noise (the so-called colored noise problem) a perturbation theory  must
be used which leads generally to an approximate description, e.g., 
within  a generalized master equation approach \cite{mukamel}. 

What to do, however, when the ambient
noise has a non-Gaussian statistics and/or it does exhibit long-range, e.g., 
power law temporal correlations with a very large, practically 
infinite range? A relevant
example is given by a $1/f^{\alpha}$ noise which is ubiquitous
in the amorphous solids and other glass-like materials like
proteins \cite{weismann,lowen}. 
Any perturbation theory in such situations
will certainly fail and we are confronting with a rather difficult problem.
Nevertheless, for an arbitrary quantum dynamics the problem of 
finding the corresponding noise-averaged
propagator can be solved exactly, at least formally, 
for a rather general class of 
non-Markovian jump processes  modeled as a continuous-time random 
walk (CTRW) \cite{montroll,lax,hughes} within arbitrary, but finite
number of states $N$. 
A similar problem has been already
considered in several previous works, notably in 
Ref. \cite{kampen79,chvosta}. 
These works did not solve, however, the problem at hand for the 
case of {\it stationary}
noise-averaging for a multi-state non-Markovian noise, 
when the evolution of the considered stochastic process starts 
in the infinite past, and not simultaneously with the
evolution of the considered dynamical system. In the case of a non-Markovian
noise this problem is {\it not} trivial. Physically this is but the
most important and relevant case to study. 

In this work we utilize the most general description of the discrete
state non-Markovian processes of the continuous time random 
walk (CTRW) type with uncorrelated jumps. Generally, such processes are defined by the set
of probability
densities $\psi_{ij}(\tau)$ for making transitions among the 
discrete noise states \cite{hughes}.
The noise-averaged quantum propagator is obtained below
for this general case. 
The approaches of Ref. \cite{kampen79} and Ref. 
\cite{chvosta} are unified within this general description. 
Next, the problem of stationary noise averaging is considered.
It is shown in a constructive way how to make the stationary noise 
averaging in the case of 
factorized probability densities like in Ref. \cite{chvosta}, but
with the time-independent matrix of transition probabilities $p_{ij}$.
In this respect, the present work is most close to Ref. \cite{burstein86},
where a similar approach has been proposed, however the explicit
solution for the stationary noise averaged quantum propagator has not been 
obtained.
The corresponding formally {\it exact}
expression for the  Laplace-transformed averaged
propagator is found in this work in the explicit form for the first time.
This presents the {\it first} main result of this work  which possesses
an ample range of applications. In particular,
the noise-averaged propagator of the Kubo-oscillator 
with a stochastically modulated frequency, as well as the corresponding
line shape form are also obtained. 
This presents the {\it second} main result of this work which
can be important, e.g., for the single-molecular spectroscopy.

\section{Model and Theory}
   
Let us consider an arbitrary quantum system with
the Hamilton operator $\hat H[\xi(t)]$ which depends parametrically
on the stochastic process $\xi(t)$ which in turn can acquire
randomly in time $N$ discrete values $\xi_i$. Accordingly,
the Hamiltonian $\hat H[\xi(t)]$ can take on 
$N$ different random operator
values $\hat H[\xi_i]$.
The discrete stochastic
process is assumed to be a non-Markovian renewal process which is 
fully characterized by
the set of  probability densities $\psi_{ij}(\tau)$
which describe random transitions among the states $\xi_i$. 
Namely, $\psi_{ij}(\tau)$
is the probability density for making transition from
the  state $j$ to the state
$i$. These probability densities must obviously be positive and obey the
normalization conditions 
\begin{equation}\label{norm}
\sum_{i=1}^{N}\int_{0}^{\infty}
\psi_{ij}(\tau)d\tau =1, 
\end{equation}
for all $j=\overline{1,N}$. 
All random transitions are assumed to be mutually 
independent \cite{remark0}.
The residence time distribution (RTD) $\psi_j(\tau)$ in the state $j$ 
reads obviously 
\begin{equation}\label{RTD}
\psi_j(\tau)=\sum_i \psi_{ij}(\tau).
\end{equation}
 The survival probability $\Phi_j(\tau)$ 
of the state $j$ follows then as 
\begin{equation}\label{survival}
\Phi_j(\tau)=\int_{\tau}^{\infty}
\psi_j(\tau)d\tau.
\end{equation}
This is the most general description used in the CTRW 
theory \cite{hughes}.

The problem is to average the quantum dynamics in the Liouville space
which is characterized by the Liouville-von-Neumann equation
\begin{equation}\label{Liouville}
\frac{d}{dt} \rho (t)= -i {\cal L}[\xi(t)]\rho(t)
\end{equation}
for the density operator $\rho(t)$ over the realizations 
of noise $\xi(t)$. ${\cal L}[\xi(t)]$ 
in Eq. (\ref{Liouville})
stands for the quantum Liouville 
superoperator, ${\cal L}[\xi(t)](\cdot)=\frac{1}{\hbar}
[\hat H[\xi(t)], (\cdot)]$. In other words, one has to find
the noise-averaged propagator 
\begin{equation}\label{propagator}
\langle U (t) \rangle = \langle {\cal T} \exp[-i\int_{t_0}^{t_0+t}
{\cal L}[\xi(\tau)] d\tau] \rangle,
\end{equation}
where ${\cal T}$ denotes the time-ordering operator.
It is the major advance of this work that we obtain the
Laplace-transform of propagator (\ref{propagator})
in the {\it exact} form for {\it arbitrary non-Markovian processes $\xi(t)$
of the discussed form}. The results of previous research
work done within the framework of SLE 
approach  for the discrete Markovian processes \cite{kubo}
follows as a particular Markovian limit
of the developed non-Markovian theory. Moreover, some previous non-Markovian
theories, notably that by van Kampen \cite{kampen79}
and that by Chvosta and Reineker \cite{chvosta} are also included
as different interpretations of this general formulation. 

In particular, the approach of van Kampen is reproduced
by introduction of the time-dependent {\it age-specific} 
rates $k_{ij}(t)$ like in the renewal theory \cite{cox}.
The probability densities $\psi_{ij}(\tau)$ then 
read \cite{kampen79,remark1}
\begin{equation}\label{kamp}
\psi_{ij}(\tau):=k_{ij}(\tau)\exp[-\sum_i\int_{0}^{\tau}k_{ij}(t)dt]. 
\end{equation}
The Markovian case corresponds to $k_{ij}(\tau)=const$.
Any deviation of $\psi_{ij}(\tau)$ from the strictly exponential
form which yields a time-dependence of the transition rates $k_{ij}(\tau)$ 
amounts to a non-Markovian behavior \cite{remark_new}. Furthermore, 
the survival probability $\Phi_j(\tau)$ in the
state $j$ within the time-dependent rate 
description is given by  
\begin{equation}\label{kamp2}
\Phi_j(\tau)=\exp[-\sum_{i=1}^{N}
\int_{0}^{\tau}k_{ij}(t)dt] 
\end{equation}
and Eq. (\ref{kamp}) then can be recast as
\begin{equation}\label{kamp3}
\psi_{ij}(\tau):=k_{ij}(\tau)\Phi_j(\tau). 
\end{equation}
The introduction of time-dependent
rates is one possible way to describe the non-Markovian effects \cite{remark1}.
It is not unique. For example,   
Chvosta and Reineker adopted a quite different and more general 
standpoint \cite{chvosta}. 
Namely, they defined
\begin{equation}\label{alter}
\psi_{ij}(\tau):=p_{ij}(\tau)\psi_j(\tau)
\end{equation}
 with $\sum_i p_{ij}(\tau)=1$.
The interpretation is as follows. The process stays 
in the state $j$ during a random time interval characterized by the
probability density $\psi_j(\tau)$. At the end of this time interval
the process makes jump into the state $i$ with a generally 
time-dependent conditional probability $p_{ij}(\tau)$.
Indeed, any stochastic process of the considered kind can be interpreted
in this way.
 For some particular applications 
in \cite{chvosta} the probability densities $\psi_j(\tau)$ 
were taken strictly exponential and all
the non-Markovian effects were assumed to come from 
the {\it time-dependent}
transition probabilities $p_{ij}(\tau)$. 
By equating Eq. (\ref{kamp3}) and Eq. (\ref{alter}) and taking into
account that $\psi_j(\tau):=-d\Phi_j(\tau)/d\tau$ it is easy to see
that van Kampen approach can be reduced to that of Chvosta and Reineker
with the time-dependent transition probabilities
\begin{equation}\label{corres}
p_{ij}(\tau)=\frac{k_{ij}(\tau)}{\sum_i k_{ij}(\tau)}
\end{equation}
and with the non-exponential probability densities $\psi_j(\tau)$ which
follow as  $\psi_j(\tau)=
\gamma_j(\tau)\exp[-\int_{0}^{\tau}\gamma_j(t)dt]$ with
$\gamma_j(\tau):=\sum_ik_{ij}(\tau)$.

The description of non-Markovian effects with the
time-dependent transition probabilities $p_{ij}(\tau)$ seems, 
however, be merely a
theoretical device. It appears to be rather difficult (if possible) 
to obtain $p_{ij}(\tau)$ from the sample trajectories of 
an experimentally {\it observed} random process $\xi(t)$.
In view of Eq. (\ref{corres}) the same is valid for
the concept of time-dependent rates. 
These rates cannot be measured directly from
the sample trajectories.
On the contrary, the RTD $\psi_j(\tau)$ and the
{\it time-independent} $p_{ij}$ (with $p_{ii}:=0$ what is assumed
in the following) 
can be routinely deduced from 
the sample trajectories measured in a {\it single-molecular} experiment. 
This latter
description is definitely more advantageous from the practical point
of view. 
From the experimentally well-defined quantities, 
$\Phi_j(\tau)$ and $p_{ij}$, the corresponding 
time-dependent rates description
can be readily found as
\begin{equation}\label{reduce}
k_{ij}(\tau)=-p_{ij}\frac{d\ln[\Phi_j(\tau)]}{d\tau}.
\end{equation}
Moreover, as it will be shown below namely this description 
with constant $p_{ij}$  in Eq. (\ref{alter}) 
does provide the consistent way in
order to construct the {\it stationary} realizations of 
the {\it non-Markovian} 
process $\xi(t)$
and, therefore, 
in order to find $\langle U(t)\rangle$ averaged, correspondingly, 
over the {\it stationary}
realizations of $\xi(t)$. These circumstances give definite advantages 
of the approach with factorized $\psi_{ij}(\tau)$ and 
{\it time-independent} $p_{ij}$ as compare with the 
formulations in \cite{kampen79} and \cite{chvosta},
even though the technical details are quite similar. 
The correct 
quantum-mechanical propagator averaged over the {\it stationary} 
realizations of a {\it non-Markovian} discrete-state noise 
with an arbitrary number of states is obtained in this work 
for the first time. This corresponds to
the situation where the quantum system was prepared at the time $t_0$
in a non-equilibrium state described by the density matrix $\rho(t_0)$ , 
but the noise was not specially prepared but it has been already 
relaxed to its stationary state. The ambient noise
evolution was started in the infinite past and one assumes that 
the initial preparation of the quantum system in a non-equilibrium state 
has no influence on the noise source. This is the most important 
physical situation to confront with.  

The task of performing the noise-averaging of the quantum dynamics 
in Eq. (\ref{propagator}) can be solved exactly due the piecewise
constant character of the noise $\xi(t)$ \cite{burshtein,frisch}. 
Namely, let us
consider the time-interval $[t_0,t]$ and to take a frozen realization
of $\xi(t)$ assuming $k$ switching events within 
this time-interval at the time-instants $t_i$, 
\begin{equation}\label{time}
t_0<t_1<t_2<...<t_k<t.
\end{equation}
Correspondingly, the noise takes on the values 
$\xi_{j_0},\xi_{j_1},...,\xi_{j_k}$ in the time sequel.
Then, the propagator $U(t,t_0)$ reads obviously
\begin{equation}\label{pathU}
U(t,t_0)=e^{-i {\cal L}[\xi_{j_k}](t-t_{k})}
e^{-i {\cal L}[\xi_{j_{k-1}}](t_k-t_{k-1})}...e^{-i {\cal L}
[\xi_{j_0}](t_1-t_{0})} 
\end{equation}
Let us assume first that it is known with certainty that at the time
instant $t_0$ the process $\xi(t)$ has {\it just started} its sojourn in
the state $j_0$. In other words, the process $\xi(t)$ has been {\it prepared}
in the state $j_0$ at $t_0$. Then, 
the corresponding $k-$times probability density
for such noise realization is 
\begin{equation}\label{prob}
P_k(\xi_{j_k},t_{k};\xi_{j_{k-1}},t_{k-1};...;\xi_{j_1},t_{1}|\xi_{j_0},t_{0})=
\Phi_{j_k}(t-t_k)\psi_{j_kj_{k-1}}(t_k-t_{k-1})...
\psi_{j_1j_0}(t_1-t_0).
\end{equation}
In order to obtain the noise-averaged $\langle U(t,t_0)\rangle_{j_0}$
conditioned on such {\it nonstationary} initial noise preparation 
one has to
average (\ref{pathU}) with the probability measure in (\ref{prob})
(for $k=\overline{0,\infty}$).
Literally (operationally) this means the following. First, 
one has to construct the time-ordered product of (\ref{pathU}) and (\ref{prob}), 
i.e.
\begin{equation}\label{product}
\Phi_{j_k}(t-t_k)e^{-i {\cal L}[\xi_{j_k}](t-t_{k})}
\psi_{j_kj_{k-1}}(t_k-t_{k-1})e^{-i {\cal L}[\xi_{j_{k-1}}](t_k-t_{k-1})}...
\psi_{j_1j_0}(t_1-t_0)e^{-i {\cal L}[\xi_{j_0}](t_1-t_{0})}.
\end{equation} 
Second, one has to perform the $k$-dimensional time integration 
of (\ref{product})
over the variables $\{t_k\}$ within the time-ordered  
domain (\ref{time}) and to sum the results over all possible $\{j_k\}$. 
Furthermore, this procedure 
has to be repeated for every $k=\overline{0,N}$ 
and the results summed at the end. 
The case $k=0$ is special with 
\begin{equation}\label{zero}
P_0(\xi_{j_0},t_0)=\Phi_{j_0}(t-t_0).
\end{equation}

The just outlined task can be easily done formally by use of the 
Laplace-transform, denoted in the following as $\tilde A(s):=
\int_{0}^{\infty}\exp(-s\tau) A(\tau) d\tau$ for any time-dependent 
quantity $A(\tau)$. For this goal, 
let us introduce two auxiliary matrix operators $\tilde A(s)$ and
$\tilde B(s)$ with the matrix elements
\begin{equation}\label{aux1}
\tilde A_{kl}(s):=\delta_{kl}
\int_{0}^{\infty}\Phi_l(\tau)e^{-(s+i {\cal L}[\xi_{l}])\tau}
d\tau
\end{equation}
and 
\begin{equation}\label{aux2}
\tilde B_{kl}(s):=
\int_{0}^{\infty}\psi_{kl}(\tau)e^{-(s+i {\cal L}[\xi_{l}])\tau}
d\tau\;,
\end{equation} 
correspondingly. Then all what remains to do is to sum the geometric
matrix operator series. The result reads,
\begin{equation}\label{result1}
\langle \tilde U(s)\rangle_{j_0}=\sum_i \Big 
(\tilde A(s)[I-\tilde B(s)]^{-1} \Big )_{i j_{0}},
\end{equation}
where $I$ is the unity matrix. It is quite obvious that instead
of the quantum Liouville operator in Eq. (\ref{Liouville}) there 
could be any linear operator, e.g. $\cal L$ be 
a matrix and $\rho(t)$ be then a
vector function.
Then, the developed theory can be immediately applied
to the averaging of arbitrary linear stochastic differential equations 
\cite{frisch,kampenBook}. Even nonlinear stochastic differential equations 
can be attempted to deal with by introducing a Liouville
equation for the corresponding classical probability density \cite{kampenBook}. 
The earlier results in \cite{kampen79} and
\cite{chvosta} are reproduced immediately from Eqs. 
(\ref{aux1})-(\ref{result1}) by an appropriate modification of the
considered problem and specifying the transition probability
densities $\psi_{ij}(\tau)$ from a most general (non-factorized form)
to a particular representation in accord with the above discussion. 

The derived result in Eqs.(\ref{aux1})-(\ref{result1}) corresponds to the 
initial preparation of $\xi(t)$ in a particular 
state $j_0$. Experimentally, this presents a quite unusual and 
strongly non-equilibrium situation. For a stationary environment
one has to perform yet an additional averaging of 
$\langle \tilde U(s)\rangle_{j_0}$ over the initial
distribution $p_{j_0}(t_0)$ taken as the 
stationary distribution, i.e.,  $p_{j_0}(t_0)=p_{j_0}^{st}$,
where $p_{j}^{st}=\lim_{t\to\infty}p_j(t)$.  Indeed, 
this presents a valid prescription for stationary noise-averaging
in the Markovian case. 
However, in a non-Markovian case 
this prescription 
is not sufficient.

Quite generally, the stationarity of noise realizations in the strict sense 
requires \cite{kampenBook} that 
not only the single-time distribution $p_j(t)$, but also 
all the multi-time joint 
probability distributions of the given process must
be invariant of a simultaneous time shift of all time arguments. However,
for many physical applications,  
the stationarity in a weak sense, i.e.,
on the level of the two-time, $P(j,t;j_0,t_0)$, joint
distribution is sufficient. Then, e.g., the stationary power
spectrum of the corresponding 
process can be defined. This two-time joint distribution 
can be expressed as  
$P(j,t;j_0,t_0)=\Pi_{jj_0}
(t|t_0)p_{j_0}(t_0)$ via the conditional probabilities 
$\Pi_{jj_0}(t|t_0)$ (propagator of the process). 
For a stationary process the consistency condition,
  $p_j^{st}=\sum_{j_0}\Pi_{jj_0}(t|t_0) p_{j_0}^{st}$, must
be satisfied for all times, i.e. $p_j^{st}$ has to be the fixed
point of the corresponding propagator (see Appendix A).
The propagator of non-Markovian process having this property
can be called quasi-stationary. 
In the present case, in order to construct such propagator and the
corresponding stationary realizations of the noise trajectories the
probability density of the {\it first} time intervals must differ from
the all subsequent ones. Indeed, 
if a noise state $j$ was occupied at $t=t_0$ with
the stationary probability $p_j^{st}$, it is not known for how
long this state was already occupied {\it before} $t_0$. The proper 
conditioning
on and averaging over this unknown time must be made and the corresponding
survival probabilities for the {\it first} residence time
interval $\tau_0=t_1-t_0$ be introduced \cite{remark3}. 
These survival probabilities read \cite{cox,remark3,remark4}
\begin{equation}\label{cond1}
\Phi^{(0)}_j(\tau)=\frac{\int_{\tau}^{\infty}\Phi_j(\tau')d\tau'}
{\langle \tau_j\rangle},
\end{equation}  
where $\langle \tau_j\rangle=\int_0^{\infty}\Phi_j(\tau)d\tau$ is
the mean residence time (MRT) of the noise in the state $j$.
Only for  strictly exponential survival probabilities, i.e., in
the Markovian case,
$\Phi^{(0)}_j(\tau)=\Phi_j(\tau)$. Otherwise, this is not the case.
The corresponding residence time distributions follow immediately
as the negative
time derivative of $\Phi_j^{(0)}(\tau)$ in Eq. (\ref{cond1}) 
and are known to be \cite{cox,lax}
\begin{equation}\label{cond2}
\psi^{(0)}_j(\tau)= \frac{\Phi_j(\tau)}
{\langle \tau_j\rangle}\;.
\end{equation}  
The {\it first time} 
transition densities 
$\psi_{ij}^{(0)}(\tau)$ then follow as
\begin{equation}\label{cond3}
\psi^{(0)}_{ij}(\tau)= p_{ij}\frac{\Phi_j(\tau)}{\langle \tau_j\rangle}.
\end{equation} 
For the logical consistency of this definition with the 
consideration pursued in  remark
\cite{remark3} $p_{ij}$ {\it must be} time-independent constants.
Otherwise, a logical problem emerges: How to make the proper conditioning 
of $p_{ij}(\tau)$ on the unknown
times before $t_0$? This is the reason why within the approaches 
of time-dependent
$p_{ij}(\tau)$, or time-dependent rates $k_{ij}(\tau)$ it is rather
obscure how
to solve the problem of stationary noise-averaging.  Therefore,
these approaches do not seem suit well for this stated purpose. 

In accord with the above discussion, the transition
density $\psi_{j_1j_0}(t_1-t_0)$ in Eq. (\ref{prob}) must be replaced
by $\psi_{j_1j_0}^{(0)}(t_1-t_0)$ from Eq. (\ref{cond3}).   
Moreover, $\Phi_{j_0}(t-t_0)$ in Eq. (\ref{zero}) must be replaced
by $\Phi_{j_0}^{(0)}(t-t_0)$ from Eq. (\ref{cond1}).
To account for these modifications, 
two auxiliary quantities $\tilde A^{(0)}_{kl}(s)$
and $\tilde B^{(0)}_{kl}(s)$ are introduced which are given by 
the expressions similar to
Eq. (\ref{aux1}) and Eq. (\ref{aux2}), but with $\Phi_{j}^{(0)}(\tau)$ instead
of $\Phi_{j}(\tau)$
and $\psi_{ij}^{(0)}(\tau)$ instead of $\psi_{ij}(\tau)$, i.e.,
\begin{equation}\label{aux1a}
\tilde A_{kl}^{(0)}(s):=\frac{\delta_{kl}}{\langle \tau_l\rangle}
\int_{0}^{\infty}e^{-(s+i {\cal L}[\xi_{l}])\tau}
 \int_{\tau}^{\infty}\Phi_l(\tau')d\tau'd\tau
\end{equation}
and 
\begin{equation}\label{aux2a}
\tilde B_{kl}^{(0)}(s):=\frac{p_{kl}}{\langle \tau_l\rangle }
\int_{0}^{\infty}\Phi_{l}(\tau)e^{-(s+i {\cal L}[\xi_{l}])\tau}
d\tau\;=\; \frac{p_{kl}}{\langle \tau_l\rangle} \tilde A_{ll}(s).
\end{equation} 
The resulting geometric operator series can again be easily
summed exactly. For the (stationary) noise-averaged Laplace-transformed 
propagator $\langle \tilde U(s)\rangle$ we  obtain:    
\begin{equation}\label{result2}
\langle \tilde U(s)\rangle=\sum_{ij} \Big 
(\tilde A^{(0)}(s)+\tilde A(s)[I-\tilde B(s)]^{-1} \tilde B^{(0)}(s)
\Big )_{i j} p_j^{st},
\end{equation}
where 
\begin{equation}\label{stat}
p_j^{st}=\frac{\langle \tau_j \rangle }{\sum_k\langle \tau_k \rangle}
\end{equation}
are the stationary occupation probabilities of $\xi_j$ (see Eq. 
(\ref{eqpop}) in Appendix A and the corresponding discussion). 
This result
can be brought into a physically more insightful form by using the
identity $\int_{\tau}^{\infty}\Phi_j(\tau')d\tau'=\langle \tau_j \rangle
-\int_{0}^{\tau}\Phi_j(\tau')d\tau'$ and upon introducing two new auxiliary
quantities
\begin{equation}\label{aux1new}
\tilde C_{kl}(s):=\delta_{kl}
\int_{0}^{\infty}e^{-(s+i {\cal L}[\xi_{l}])\tau}
 \int_{0}^{\tau}\Phi_l(\tau')d\tau'd\tau
\end{equation}
and 
\begin{equation}\label{aux2new}
\tilde D_{kl}(s):=\delta_{kl}
\int_{0}^{\infty}\psi_l(\tau)e^{-(s+i {\cal L}[\xi_{l}])\tau}
 d\tau \;.
\end{equation}  
Finally we obtain
\begin{equation}\label{final}
\langle \tilde U(s)\rangle=\langle \tilde U(s)\rangle_{static}
-\frac{1}{T}\sum_{ij} \Big 
(\tilde C(s)-\tilde A(s)[I-P\tilde D(s)]^{-1} P\tilde A(s)
\Big )_{i j},
\end{equation}
where $\langle \tilde U(s)\rangle_{static}$ is the Laplace-transform of 
the statically averaged propagator 
\begin{equation}\label{static}
\langle U(\tau)\rangle_{static}:=\sum_k e^{-i {\cal L}[\xi_{k}]\tau}p_k^{st}, 
\end{equation}
$T=\sum_k\langle \tau_k \rangle$ is the sum of mean residence times,
and $P$ is the matrix of transition
probabilities $p_{ij}$ (``scattering matrix'' of the random process 
$\xi(t)$). 
The result
in Eqs. (\ref{final}), (\ref{static}) together with Eqs. (\ref{aux1}), 
(\ref{aux1new}), (\ref{aux2new}),  presents the cornerstone
result of this work which can be used in numerous applications.

\section{Application: Kubo-oscillator}

As a simplest practical example
we consider the averaging of the so-called Kubo-oscillator 
\begin{equation}\label{kubo-osc}
\dot x(t)=i\omega[\xi(t)] x(t)\;.
\end{equation}
This particular problem appears in the theory of optical line shapes, 
in the nuclear magnetic resonance \cite{kubo,anderson53}, and in 
the single molecular 
spectroscopy \cite{jbs02}. In Eq. (\ref{kubo-osc}),  
$\omega[\xi(t)]$ presents a stochastically modulated 
frequency of quantum 
transitions between the levels of a ``two-state atom'', or between the
eigenstates of a spin 1/2 which are caused by the action of a resonant 
laser, or magnetic field, respectively.
 The spectral line shape is determined through
the corresponding stochastically averaged propagator of Kubo-oscillator 
as \cite{kubo}
\begin{equation}\label{shape}
I (\omega)=\frac{1}{\pi}\lim_{\epsilon \to +0} {\rm Re}
[ \tilde U(i\omega+\epsilon)]\;.
\end{equation}
Note that the limit $\epsilon\to +0$ in Eq. (\ref{shape}) is necessary
for the regularization of the corresponding integral in the quasi-static 
limit $T\to \infty$. 
By identifying ${\cal L}[\xi_k]$ with $-\omega_{k}$ in Eq. (\ref{final})
we obtain after some algebra
\begin{eqnarray}\label{propN}
\langle \tilde U(s) \rangle & = & \sum_{k}\frac{p_k^{st}}{s-i\omega_k}-
\frac{1}{\sum_k\langle \tau_k\rangle }
\sum_{k}\frac{1-\tilde \psi_k(s-i\omega_k)}{(s-i\omega_k)^2}\\ \nonumber
& + &
\frac{1}{\sum_k\langle \tau_k\rangle }
\sum_{n,l,m}\frac{1-\tilde \psi_l(s-i\omega_l)}{s-i\omega_l}
\Big (\frac{1}{I-P\tilde D(s)}\Big )_{lm}p_{mn}
\frac{1-\tilde \psi_n(s-i\omega_n)}
{s-i\omega_n}
\end{eqnarray}
with $\tilde D_{nm}(s)=\delta_{nm}\tilde \psi_m(s-i\omega_m)$. 
The corresponding
line shape follow immediately from Eq. (\ref{propN}) by virtue of 
Eq. (\ref{shape}). This result presents a non-Markovian 
generalization of the earlier result by Kubo \cite{kubo} for 
arbitrary $N$-state
discrete Markovian processes. The generalization  
consists in allowing for
arbitrary non-exponential RTDs $\psi_k(\tau)$, or, equivalently, in 
accordance with Eq. (\ref{reduce}) for time-dependent transition
rates $k_{ij}(\tau)$. This generalization is obtained here for the
first time and presents one of our main results. Let us further
simplify the result in Eq. (\ref{propN}) for the case of two-state
non-Markovian noise with $p_{12}=p_{21}=1$. Then, Eq. (\ref{propN})
yields after some simplifications:
\begin{eqnarray}\label{prop2}
\langle \tilde U(s)\rangle & = & \sum_{k=1,2}\frac{1}{s-i\omega_k}
\frac{\langle \tau_k \rangle}{\langle \tau_1 \rangle+
\langle \tau_2 \rangle}\nonumber \\
& + &\frac{(\omega_1-\omega_2)^2}{(\langle \tau_1 \rangle+
\langle \tau_2 \rangle)(s-i\omega_1)^2(s-i\omega_2)^2}
\frac{[1-\tilde \psi_1(s-i\omega_1)][1-\tilde \psi_2(s-i\omega_2)]}
{1-\tilde \psi_1(s-i\omega_1)\tilde \psi_2(s-i\omega_2)}\; .
\end{eqnarray}
 With (\ref{prop2}) in (\ref{shape}) one obtains the result 
 for the corresponding 
spectral line shape which is equivalent to one obtained  
recently in Ref. \cite{jung}
using a different method. It is reproduced here as a simplest  
application of our more general approach.

\section{Summary}

In this work the problem of the stochastic averaging of a 
quantum dynamics with 
non-Markovian fluctuating parameters has been investigated within the
trajectory description of continuous time random walk theory.
The formally exact expression for the stochastically 
averaged quantum-mechanical
propagator is obtained for the most general 
CTRW with uncorrelated jumps and for a non-equilibrium noise preparation. 
The problem of stationary noise averaging has been solved
for the practically relevant formulation 
with the time-independent matrix of transition
probabilities $p_{ij}$. Especially, 
the formally exact expression  for the stationary averaged quantum
propagator has been found in an explicit form. 
This general expression has been used on order to find
the stationary propagator of the Kubo oscillator describing the spectral
line shape of optical transitions in a two-state atom. 
Further applications, 
such as decoherence of a two-state quantum dynamics driven by
two-state non-Markovian noises, including $1/f^{\alpha}$ noise
case, are in progress.

\acknowledgements  This work has been supported by the Deutsche
Forschungsgemeinschaft via the Sonderforschungsbereich SFB-486,
{\em Manipulation of matter on the nanoscale}, project No. A10.

\appendix
\section{Propagators and generalized master equations} 

In this Appendix, both the nonstationary propagator and the  
propagator for quasi-stationary initial preparations 
for the considered non-Markovian processes are obtained,
along with the corresponding generalized master equations (GMEs).
The propagator $\Pi_{ij}$ of the process $\xi(t)$, or the matrix of 
conditional
probabilities connects the initial probability vector $\vec p(t_0)$
with the final one, $\vec p(t_0+\tau)$, i.e., 
\begin{eqnarray}\label{a1}
p_i(t_0+\tau)=\sum_{j}\Pi_{ij}(t_0+\tau|t_0)p_j(t_0).
\end{eqnarray}
The expression for the Laplace-transform
$\tilde \Pi_{ij}(s)$ can be obtain in a way similar to the averaging 
of quantum propagator in Eqs. (\ref{result1}) and (\ref{result2}). 
Basically, one has to put there
${\cal L}\to 0$. For the nonstationary propagator of $\xi(t)$, i.e., when
the process $\xi(t)$ {\it starts} its evolution 
at $t=t_0$ in a particular
state the result reads for the general case
of non-factorized probability densities $\psi_{ij}(\tau)$ as follows: 
\begin{eqnarray}\label{a2}
\tilde \Pi_{ij}(s)=\tilde \Phi_i(s)
\left ([I-\tilde \Psi(s)]^{-1} \right)_{ij},
\end{eqnarray}
where   $\Psi$ is the matrix of $\tilde \psi_{ij}(s)$ and
$\tilde \Phi_j(s)=[1-\tilde\psi_j(s)]/s$, $\tilde\psi_j(s)=
\sum_i \tilde \psi_{ij}(s)$. For quasi-stationary
initial preparations the propagator of non-Markovian process
is generally different \cite{hanggi,goychuk03}.
In the factorized
case $\psi_{ij}(\tau)=p_{ij}\psi_j(\tau)$
it reads  for the considered process (cf. Eq. (\ref{result2})):
\begin{eqnarray}\label{a3}
\tilde \Pi_{ij}^{st}(s)=\tilde \Phi_i^{(0)}(s)\delta_{ij}+
\tilde \Phi_i(s)\sum_k
\left ([I-\tilde \Psi(s)]^{-1}\right)_{ik}p_{kj}\frac{\tilde \Phi_j(s)}
{\langle \tau_j\rangle},
\end{eqnarray}
where $\tilde\Phi^{(0)}_j(s)=1/s-[1-\tilde\psi_j(s)]/
(s^2\langle \tau_j\rangle)$. The stationary populations
follow as $p_i^{st}=\lim_{s\to 0}(s\tilde\Pi_{ij}(s))$.
The quasi-stationary propagator $\Pi_{ij}^{st}(\tau|0)$ must satisfy
the consistency condition 
 $p_i^{st}=\sum_{j}\Pi_{ij}^{st}(\tau|0)p_j^{st}$ for all 
times $\tau$,
i.e., $\vec p^{st}$ is the fixed point of $\Pi_{ij}^{st}(\tau|0)$.
Let us prove this fact and to find $p_i^{st}$.
It is more convenient to do both tasks by finding first 
the corresponding GMEs for
$p_i(t)$. These generalized master equations are of substantial 
interest {\it per se}.

In order to find the corresponding GMEs the procedure 
of \cite{burstein86} can be applied
to a more general present case of non-factorized $\psi_{ij}(\tau)$. 
Indeed, let us consider the conditional probability $P_{k}(j|j_0)(t)$ 
for making $k$-jumps within the time interval $[0,t]$ starting at $t=0$
in the state $j_0$ with the probability $p_{j_0}(0)$ and finishing in
the state $j$ with the probability $p_j(t)$. This probability is given by  
a corresponding $k$-dimensional integral of 
(\ref{prob})) (see the discussion below Eq. (\ref{prob})) with the summation
made over $j_{k-1},j_{k-2},...,j_1$. The corresponding  
 Laplace-transform $\tilde P^{(k)}_{jj_0}(s)$ reads:
\begin{eqnarray}\label{probK}
\tilde P^{(k)}_{jj_0}(s)=\sum_{j_{k-1}}...\sum_{j_{1}}
\tilde \Phi_{j} (s)\tilde \psi_{jj_{k-1}}(s)...
\tilde \psi_{j_{2}j_{1}}(s)
\tilde \psi_{j_{1}j_{0}}^{(0)}(s)
\end{eqnarray}
for $k>0$.
Here, Eq. (\ref{prob}) was used in a slightly modified form with
$\psi_{ij}^{(0)}(\tau)$ for the probability densities of the first time
intervals. Furthermore, for $k=0$,
$\tilde P^{(0)}_{jj_0}(s)=\tilde \Phi_{j}^{(0)}(s)\delta_{jj_0}$,
where $\tilde \Phi_{j}^{(0)}(s)$ is the Laplace-transform of the
corresponding survival probability $\Phi_{j}^{(0)}(\tau)=\sum_i\int_{\tau}^
{\infty}\psi_{ij}^{(0)}(\tau')d\tau'$ of the first time-interval.
For $k\geq 2$,
The quantities $\tilde P^{(k)}_{jj_0}(s)$ satisfy obviously the following 
recurrence relation:
\begin{eqnarray}\label{rec1}
\tilde P^{(k)}_{jj_0}(s)=\tilde \Phi_{j} (s)\sum_{n}
\tilde \psi_{jn}(s)\frac{\tilde P^{(k-1)}_{nj_0}(s)}{\tilde \Phi_{n} (s)}\;.
\end{eqnarray}
Furthermore, the Laplace-transform of propagator $\tilde \Pi_{jj_0}(s)$
is expressed in terms of $\tilde P^{(k)}_{jj_0}(s)$ as 
$\tilde \Pi_{jj_0}(s)=\sum_{k=0}^{\infty}\tilde P^{(k)}_{jj_0}(s)$.
Then, by virtue of Laplace-transformed Eq.(\ref{a1}) with $t_0=0$,
\begin{eqnarray}\label{aux}
\sum_{k=2}^{\infty}\sum_{j_0}\tilde P^{(k)}_{jj_0}(s)p_{j_0}(0)&=& 
\tilde p_j(s)-\sum_{j_0}\tilde P^{(1)}_{jj_0}(s)p_{j_0}(0)-
\tilde P^{(0)}_{jj}(s)p_{j}(0)\nonumber \\
&=&\tilde \Phi_{j} (s)\sum_{n}
\tilde \psi_{jn}(s)\frac{1}{\tilde \Phi_{n} (s)}
\sum_{k=2}^{\infty}\sum_{j_0}\tilde P^{(k-1)}_{nj_0}(s)p_{j_0}(0)
\end{eqnarray}
where the recurrence relation (\ref{rec1}) has been used. The use
of $\sum_{k=2}^{\infty}\sum_{j_0}\tilde P^{(k-1)}_{nj_0}(s)p_{j_0}(0)=
\tilde p_n(s)-\tilde \Phi_n^{(0)}(s)p_n(0)$ in Eq. (\ref{aux})
finally yields
\begin{eqnarray}\label{inter}
\tilde p_j(s)= \tilde \Phi_j(s)\sum_n\tilde \psi_{jn}(s)
\frac{\tilde p_n(s)}{\tilde \Phi_n(s)}+\tilde \Phi^{(0)}_j(s)p_j(0)\nonumber\\
+ \tilde\Phi_j(s)\sum_n\left( \tilde \psi_{jn}^{(0)}(s)-\tilde \psi_{jn}(s)
\frac{\tilde \Phi_n^{(0)}(s)}{\tilde \Phi_n(s)}\right)p_n(0)\;.
\end{eqnarray}
Let us consider now the case when the noise has been prepared at $t_0=0$ in a
particular state with the probability one. 
Then, $\psi_{ij}^{(0)}(\tau)\equiv\psi_{ij}(\tau)$
and the last term in (\ref{inter}) vanishes. For this class of non-equilibrium 
initial
preparations, the inversion of (\ref{inter}) yields:
\begin{eqnarray}\label{GME1}
\dot p_j(t)=-\sum_n\int_0^{t}\Gamma_{nj}(t-t')p_j(t')dt'+\sum_n
\int_0^{t}\Gamma_{jn}(t-t')p_n(t')dt'
\end{eqnarray}
where the Laplace-transformed kernels 
reads
\begin{eqnarray}\label{kernels}
\tilde \Gamma_{jn}(s)=\frac{s\tilde\psi_{jn}(s)}{1-\tilde\psi_{n}(s)}
\end{eqnarray}
with $\tilde\psi_{n}(s)=\sum_j\tilde\psi_{jn}(s)$.
The just derived GME (\ref{GME1}),(\ref{kernels}) is the most general
GME for the continuous time random walk processes with uncorrelated jumps
for the given class of initial preparations.
In the case of factorized (but still non-separable) 
CTRW with $\psi_{ij}(\tau)=p_{ij}\psi_j(\tau)$
it reduces to the GME of Burstein,
Zharikov and Temkin \cite{burstein86}. 
Moreover, in assumption that all $\psi_j(\tau)$ are equal (separable
CTRW of Montroll and Weiss), 
it reduces further to the  GME of Kenkre, Montroll and
Shlesinger \cite{kenkre}. 
The stationary populations can be obtained
from (\ref{inter}) as $p_j^{st}=\lim_{s\to 0}(s\tilde p_j(s))$.
Assuming that the mean residence times $\langle \tau_j\rangle$ exist,
i.e. $\tilde \psi_{ij}(s)=\alpha_{ij}-s t_{ij}+o(s)$ with 
$\sum_{i}\alpha_{ij}=1$ and $\sum_{i} t_{ij}=\langle \tau_j\rangle$,
Eq. (\ref{inter}) yields the system of linear algebraic equations 
for the stationary populations 
\begin{eqnarray}\label{eqpop}
\frac{p_j^{st}}{\langle \tau_j\rangle}=\sum_n \alpha_{jn}\frac{p_n^{st}}
{\langle \tau_n\rangle}
\end{eqnarray}
For an ergodic process the stationary probability to find the process
in a particular state should
be proportional to the time which the process spends in this particular
state on average, i.e., be given by Eq. (\ref{stat}). It is easy
to verify that Eq. (\ref{stat}) provides the  solution 
of (\ref{eqpop}) for $\sum_{j}\alpha_{ij}=1$ (for the factorized case
$\alpha_{ij}$ coincide obviously with $p_{ij}$). This latter condition
can be also expressed as
\begin{eqnarray}\label{ergod}
\sum_{j=1}^{N}\int_{0}^{\infty}
\psi_{ij}(\tau)d\tau =1 \;. 
\end{eqnarray}

For factorized case $\psi_{jn}(\tau)=p_{jn}\psi_n(\tau)$ 
with $\psi_{j}^{(0)}(\tau)=\Phi_j(\tau)/\langle \tau_j
\rangle$ \cite{remark3}, the inversion of (\ref{inter}) yields
the GME for the discussed non-Markovian process with quasi-stationary
initial preparations:
\begin{eqnarray}\label{GME2}
\dot p_j (t)=-\int_{0}^{t}\Gamma_j(t-t')[p_j(t')-p_j(0)]dt'+
\sum_n p_{jn}\int_{0}^{t} \Gamma_n(t-t')[p_n(t')-p_n(0)]dt'\nonumber \\
-\frac{p_j(0)}{\langle \tau_j\rangle}
+\sum_n p_{jn}\frac{p_n(0)}{\langle \tau_n\rangle}\;
\end{eqnarray}
with the kernels given by 
$\tilde \Gamma_j(s)=s\tilde \psi_j(s)/(1-\tilde \psi_j(s))$.
By choosing $p_n(0)=p_{n}^{st}$ which satisfy Eq. (\ref{eqpop}) it becomes
obvious that $p_n(t)=p_n^{st}$ is the solution of (\ref{GME2})
for {\it all} times $t>0$. This means that the stationary $\vec p^{st}$ 
provides indeed the fixed
point of the corresponding quasi-stationary propagator.


\begin{references}
\bibitem{remark0} This is the so-called semi-Markov assumption.
The term ``semi-Markov'', frequently used in the mathematical
literature is, however, somewhat unfortunate. It comes from
a generalization of the discrete time Markovian chain processes to
the continuous time case. Such continuous time processes
are, however, generally non-Markovian since the statistical independence 
of the residence
time intervals does not imply  yet that all the multi-times probability
densities for the considered process can be expressed merely through
the corresponding two-times and single-time probability densities.
\bibitem{remark1}
Note that such time-dependence of $k_{ij}(\tau)$ has nothing to
do with the possible nonstationarity invoked, e.g., due to the action 
of external 
fields like in the problem of Stochastic Resonance (SR) \cite{SR}. 
Introducing such
time-dependent rates is merely
a language to describe the non-Markovian  memory effects. 
In the case of a nonstationary background noise driven by
a time-dependent signal (like in the non-Markovian 
SR problem, cf. \cite{goychuk03}) 
the use of such language in a non-Markovian
situation should be avoided
as it can potentially confuse the readers.
\bibitem{remark_new}This observation can be rationalized as follows.
Let us consider a sojourn in the state $j$
characterized by the survival
probability $\Phi_j(\tau)$. The
corresponding
residence time interval $[0,\tau]$ can be arbitrarily
divided into two pieces
$[0,\tau_1]$
and $[\tau_1,\tau]$.
If no memory effects are present, then
$\Phi_j(\tau)=\Phi_j(\tau-\tau_1)\Phi_j(\tau_1)$.
The only nontrivial solution of this latter functional equation which decays
in time reads $\Phi_j(\tau)=\exp(-\gamma_j\tau)$, with $\gamma_j>0$.
This corresponds to a Markovian case.
\bibitem{remark4} Note that no averaging over the unknown times 
$\tau^{*}$ \cite{remark3}
was made in \cite{kampen79} (see Appendix C there). 
Therefore, the problem of stationary noise-averaging was not solved
in this important paper.
\end{references}
\end{document}